\documentclass[format=manuscript, review=false, anonymous=false, screen, dvipsnames]{lib-acm/acmart}

\setcopyright{cc}
\setcctype{by}
\acmJournal{TOCE}
\acmYear{2026} \acmVolume{1} \acmNumber{1} \acmArticle{}
\acmMonth{1} \acmDOI{10.1145/3834566}

\usepackage{booktabs} % For formal tables
\usepackage{multirow}

\usepackage{exii-macros}

\usepackage{booktabs}

\showrevisions{REVISIONGREEN}

\makeatletter
\g@addto@macro\normalsize{%
  \setlength\abovedisplayshortskip{-9pt}
  \setlength\belowdisplayshortskip{3pt}
}
\makeatother

\begin{document}

% this seems to help remove words going beyond the margin
\tolerance=400 

% The title should be short but descriptive, like a mini abstract. 
% If possible, come up with a catchy name for your project and use it as part of the title.
% can include a short version of the title for the running header (in square brackets)
\title[Make or Take: How Students Navigate Self-Created and Instructor-Provided Cheat Sheets]{Make or Take: How Students Navigate Self-Created and Instructor-Provided Cheat Sheets}
% \titlenote{Produces the permission block, and
% \subtitle{Extended Abstract}
% \subtitlenote{The full version of the author's guide is available as
%   \texttt{acmart.pdf} document}


\author{Helen Weixu Chen}
\orcid{0009-0008-1384-6781}
\affiliation{%
  \institution{Cheriton School of Computer Science, 
  University of Waterloo}
  \country{Waterloo, ON, Canada}
}
\email{w352chen@uwaterloo.ca}

\author{Victoria Sakhnini}
\email{vsakhnini@uwaterloo.ca}
\affiliation{%
  \institution{Cheriton School of Computer Science, 
  University of Waterloo}
  \country{Waterloo, ON, Canada}
}
\orcid{0000-0001-6350-3885}

\author{Lesley Istead}
\orcid{0000-0003-0063-8154}
\affiliation{
\institution{School of Information Technology, Carleton University}
\country{Ottawa, ON, Canada}
}
\email{LesleyIstead@cunet.carleton.ca}

\renewcommand{\shortauthors}{Chen et al.}

\begin{abstract}
The use of cheat sheets in exams is often framed as a way to reduce cognitive load and support student performance. However, little is known about how students choose between self-created and instructor-provided cheat sheets, or how these choices relate to their broader approaches to exam preparation. We conducted a longitudinal study in a senior-level undergraduate software requirements course, where students could use either an instructor-provided or a self-created cheat sheet for both the midterm and final exams. 
Across three survey waves, we received 53, 50, and 44 responses, respectively. 41 students completed all three surveys and formed the longitudinal cohort used to examine how choices and experiences evolved over time, while exam-specific analyses used all available responses from the corresponding wave.
Our findings identify several considerations that shaped students’ choices, including trust in instructor expertise, the desire for personalization, and preparation efficiency. We further show how students’ attitudes shifted over time and how their preferences were reflected in patterns of cheat sheet use, perceived content coverage, and challenges encountered during the exams.

\end{abstract}



% keep keywords to one line in rendered paper, try to use big topics that aren't
% already in your title

% % optional full width teaser figure
%   \includegraphics[width=\textwidth, height=4cm]{figures/tbd.pdf}
%   \caption{This is a teaser}
%   \Description{This is the teaser description for screen readers.}

\maketitle

% --------------------------------
% BODY
% edit this file to insert your sections
% Exii Standard Section Index
% ================================

% sections are each in separate files

\section{Introduction}
In computing education, cheat sheets are presented as tools to reduce cognitive load and support student performance on exams \cite{erbe2007reducing, de2012student, Whitehouse02122025}. However, there exists a pedagogical split: are cheat sheets best designed by instructors to ensure consistency and fairness, or should students construct their own as a form of personalized learning? A large portion of the literature assumes that this choice is made for students by instructors or course policies \cite{danielian2019support, smith2019instructor}. What happens when students are given both options and asked to choose remains under-explored.

Previous work investigates student behaviour under fixed cheat sheet policies. Some studies examine the layout and content of self-created cheat sheets, correlating quality with exam performance \cite{song2015quantitative, de2012student, wunsche2025characteristics}. Others look to standardized instructor-provided sheets as a way to promote fairness and reduce cognitive overload \cite{danielian2019support}. Still others compare cheat sheets to open-book conditions, showing that students using them study more, but may experience more anxiety if the sheet is poorly designed \cite{gharib2012test}. The cheat sheet is handled in each of these pieces as a fixed intervention, rarely an object of student choice, negotiation, or meaning-making.

In this work, we explore how students reason about the role of cheat sheets when given a choice. Our study was conducted in a senior-level undergraduate course on software requirements. Students were offered two formats: an instructor-provided cheat sheet and the option to create their own. Each student could decide which to use for the midterm and final exam. We ask:

\begin{itemize}
\item \textbf{RQ1}: What factors shape students’ preferences between instructor-provided and self-created cheat sheets?
\item \textbf{RQ2}: How do students’ choices of cheat sheet relate to their performance, preparation strategies, and usage patterns?
\end{itemize}

To answer these questions, we conducted a longitudinal survey study across three waves: before the midterm, after the midterm and before the final exam, and after the final exam. The surveys combined closed-ended and open-ended questions about students’ cheat sheet choices, preparation strategies, and exam experiences. Rather than presupposing which format was better, we sought to understand how students interpreted the purpose, limitations, and pedagogical signals embedded in each option. We found that students’ cheat sheet decisions are shaped by three recurring tensions: 1) trust in instructor judgment versus trust in their own knowledge needs, 2) coverage versus clarity, and 3) effort versus payoff. These tensions evolve over time. Early decisions based on convenience or deference often give way to more deliberate preparation strategies, including how frequently students refer to their cheat sheets and how they adjust their format for greater effectiveness. In these acts of preparation, we see not just study behaviour, but an attempt to reconcile what students are expected to know with how they feel supported in knowing it.

In the discussion, we consider how cheat sheet policy might be reframed, not as a technical constraint, but as a pedagogical design choice that communicates expectations, grants or withholds agency, and co-defines the learning contract between student and instructor.

\section{Background and Related Work}
\subsection{Student-Created Testing Aids}
Student-prepared notes permitted during exams have long served as a pedagogical device for active learning. Rooted in the belief that making the aid deepens understanding, cheat-sheet preparation engages students in processes of selection, organization, and synthesis \cite{larwin2012student, erbe2007reducing}. Across decades of educational research, this intuition has been supported by findings that connect the act of note construction to deeper cognitive encoding and meta-cognitive awareness \cite{larwin2012student, de2012student}. Yet, as studies across domains reveal, the benefits of this practice are far from uniform. In programming and economics courses alike, students who crafted their own sheets tended to perform better on conceptual and analytical questions \cite{Settlage02012019}, suggesting that the value of the cheat sheet lies in how it externalizes understanding rather than reproduces content mechanically \cite{de2012student, wunsche2025characteristics}. Sheets emphasizing abstract representations, such as pseudocode, formulas, or design principles, predict higher learning gains than those densely filled with worked examples or copied code \cite{larwin2012student, de2012student, wunsche2025characteristics}. Coverage also matters; incomplete sheets were associated with poorer outcomes \cite{wunsche2025characteristics}.

However, this optimistic picture is not universal. Several psychology studies report negligible or null effects on exam performance \cite{hindman1980crib, dickson2005authorized}. \citet{funk2011crib} even observed that students expecting to use cheat sheets but later denied them performed worse than those who never expected one, an effect attributed to over-reliance and misplaced confidence. Moreover, the preparation of such aids imposes a non-trivial time cost \cite{Whitehouse11122024}, raising concerns about equity: students with fewer resources, limited time, or weaker organizational skills may struggle to produce equally effective sheets.

Taken together, prior research treats cheat-sheet creation as a universal expectation and investigates its cognitive or content-based correlates. However, this literature overlooks a crucial question: what happens when students are given a choice to create their own aid or to adopt a pre-made one? To the best of our knowledge, no existing studies examine how learners reason about this trade-off between effort and benefit, nor how differences in self-efficacy or available time shape that decision. Our work addressed this gap by reframing the cheat-sheet task as a site of choice, reflection, and resource negotiation rather than a uniform instructional practice.

\subsection{Instructor-Provided Exam Materials}
Previous work explored instructor-provided exam materials, formula sheets, reference lists, and standardized handbooks, as a means of reducing exam anxiety and ensuring fairness. These resources offered consistent access to key information and removed disparities in preparation effort. Common examples include mathematics formula sheets \cite{song2015quantitative} and standardized reference manuals used in professional engineering exams \cite{smith2019instructor}. 

Early studies, however, revealed that instructor-provided materials had limited benefits. \citet{dickson2005authorized} found no significant improvement in performance or stress reduction when support sheets were distributed by instructors, in contrast to later findings where self-created sheets led to measurable gains. This difference points to a key distinction: when information is supplied rather than constructed, learners engage with it more passively. Similar trends emerged in earlier open-book exam research \cite{boniface1985candidates, feldhusen1961evaluation}, where students relying on provided resources prepared less intensively, anticipating that external aids could compensate for limited study.
These findings align with long-standing educational theories emphasizing the distinction between active and passive cognitive processing. Educational theorists have long argued that such passivity limits meta-cognitive engagement, while self-prepared aids encourage learners to identify gaps and synthesize understanding \cite{erbe2007reducing, larwin2012student}. 

Recent work refines instructor involvement. \citet{song2015quantitative} caution that distributing ``sample cheat sheets'' undermines active learning benefits, recommending instead that instructors provide scaffolds—organizational frameworks that support student strategy without replacing effort.  
In professional and certification settings, however, the rationale differs: standardized reference manuals, such as the 160-page Fundamentals of Engineering handbook, aim to simulate authentic workplace practice rather than reinforce conceptual learning \cite{smith2019instructor}. 

Despite this rich body of work comparing the two, little attention has been given to the decision space between them: how students reason about whether to rely on instructor-provided materials or invest effort in creating their own. Our work explored how students weigh efficiency against personalization, authority against autonomy, and cognitive benefit against time cost when both options coexist.

\subsection{Cheat Sheets in Software Requirements Engineering Education}
Most studies on cheat sheets in computing education focus on implementation-oriented contexts such as introductory programming \cite{de2012student}, data structures \cite{hamouda2016crib}, and computer graphics \cite{wunsche2025characteristics}. These works examine how students summarize syntax, algorithms, or formulas to aid code writing and debugging. Yet such findings may not generalize to conceptually oriented domains like software requirements engineering, where the emphasis shifts from implementation to reasoning, modeling, and communication.

Requirements engineering is a design-centered discipline that asks students to interpret stakeholder needs and represent them through user stories, UML diagrams, or formal specifications \cite{Norheim_Rebentisch_Xiao_Draeger_Kerbrat_de_Weck_2024}. Success depends less on recalling code patterns and more on synthesizing perspectives, managing ambiguity, and choosing appropriate modeling techniques \cite{berry2012case}.
The domain’s interdisciplinary nature also blends analytical and communicative skills, involving stakeholder dialogue \cite{boehm1998stakeholder}, architectural reasoning \cite{tang2007rationale}, and formal notation \cite{sitaraman2001formal}. These forms of reasoning engage different cognitive processes, suggesting that the strategies students use to prepare cheat sheets may differ from those observed in code-intensive courses.

Moreover, the learner population differs. Prior studies mainly involve novices who benefit from externalized memory support. In contrast, our participants are senior students with extensive programming experience and more developed meta-cognitive skills. For them, cheat sheet preparation may be a strategic act—balancing personalization with efficiency and leveraging organizational experience to optimize exam performance. How such advanced learners reason about or benefit from cheat sheets in conceptual, interdisciplinary settings remains an open question.

\section{Method}
We engaged students through three surveys distributed throughout the term using the course platform’s built-in survey feature. Two authors managed survey setup and communication with the class. Responses were downloaded and cleaned by one author to prepare them for analysis. All authors participated in the coding process and contributed to discussions around code development and refinement. 

\subsection{Procedure and Analysis}
To capture students' experiences around the midterm and final exam, we administered three surveys to students enrolled in the Winter 2025 offering: Survey 1 before the midterm, Survey 2 after the midterm, and Survey 3 after the final exam. Each survey included multiple-choice and open-ended questions. We received 53, 50, and 44 responses, respectively. 
% Because our analysis focused on patterns across all three time points, we included the 41 students who completed all three surveys.
Analyses of responses from a single survey wave used all available responses from that wave. Thus, analyses of initial preferences used up to 53 responses from Survey 1, analyses of midterm experiences used up to 50 responses from Survey 2, and analyses of final exam experiences used up to 44 responses from Survey 3. Analyses examining changes in students' choices or attitudes across the term were restricted to the 41 students who completed all three surveys. 

Surveys were designed to capture students' cheat sheet format preferences, rationales, and usage patterns. The full list of survey questions can be found in Appendix \ref{apx:survey questions}. To be specific, students were asked which cheat sheet format they chose, why they made that choice, whether they would choose differently next time, and how they modified their cheat sheet if they created one themselves, or what changes they would have suggested if they used the instructor-provided version. Beyond these preference and rationale questions, surveys also investigated students’ perceptions of cheat sheet coverage, their strategies for referencing the cheat sheet during exams, and the time they spent on exam preparation. 

We approached the open-ended responses through a combination of deductive and inductive coding \cite{deterding2021flexible}. While initial coding categories were informed by prior literature on exam preparation (e.g., anxiety \cite{erbe2007reducing}, effort \cite{de2012student}, trust in instructional material \cite{gharib2012cheat}), we remained open to emergent themes, such as personalization, self-efficacy, and perceived alignment with the exam that arose through close reading. One author led the initial round of thematic grouping \cite{braun2021one, braun2012thematic}, after which the team collaboratively refined codes, resolved ambiguities, and synthesized cross-cutting patterns. 

\subsection{Recruitment and Participants}
A total of 55 students were enrolled in the Winter 2025 offering of the course. Students were invited to participate in the study in exchange for up to 5 bonus marks. Participation was voluntary; those who chose not to participate could instead complete a 500-word written reflection for equivalent credit. Recruitment occurred through course announcements, and students consented separately at each stage of the study. The 5 bonus marks were distributed across the three surveys: Survey 1 and Survey 2 were each worth 1 mark, while the longer Survey 3 was worth 3 marks. Students who completed only some of the surveys would receive partial marks accordingly.

Importantly, our study did not assign students to experimental conditions. All students were presented with two options for both the midterm and final exams: they could use an instructor-provided cheat sheet or prepare their own, subject to a maximum of one double-sided page. The instructor did not disclose any preference or recommendation regarding which option to select, and the instructor-provided cheat sheet was intentionally not shown to students prior to the exam. This design prevented the instructor’s version from influencing participants’ decisions or preparation strategies, allowing us to observe student preferences in a naturalistic yet controlled context across multiple high-stakes assessments.

Within the longitudinal cohort of 41 participants who completed all three surveys, 20 were men, 19 were women, 1 identified as questioning, and 1 preferred not to disclose their gender. Over 90\% indicated prior experience using cheat sheets in other courses.

\subsection{Ethics and Privacy}
This study was approved by the relevant research ethics board. 
Although responses could be linked across survey rounds through the course platform to identify participants who completed all three surveys, the data were de-identified prior to analysis. During de-identification, participants were assigned participant IDs (e.g., P1, P2). 
These IDs were retained throughout analysis to allow consistent reference across survey rounds and qualitative excerpts, but they were not re-numbered after excluding incomplete cases. 
% As a result, participant IDs referenced in the paper do not necessarily range from P1 to P41, even though our analysis focused on the 41 students who completed all three surveys. 
No directly identifying information was included in the dataset used for coding and interpretation, beyond any optional self-disclosures participants chose to include in open-text responses.

\renewcommand{\arraystretch}{1.4}

\begin{table*}[t]
\centering
\caption{Overview of key themes with selected representative quotes from participants}
\label{tab:qual_themes}
\begin{tabular}{|p{3cm}|p{5cm}|p{8cm}|}
\hline
\textbf{Category} & \textbf{Theme} & \textbf{Representative Quote} \\
\hline
\multirow{6}{=}{\textbf{Rationales Behind Choices}} 
& Trust in Instructor & \textit{“I trust the instructor would make it useful.”} (P20) \\
\cline{2-3}
& Cognitive Efficiency & \textit{“Less pressure on making a cheat sheet.”} (P17) \\
\cline{2-3}
& Exam Alignment & \textit{“The instructor will include more relevant information to the exam.”} (P1) \\
\cline{2-3}
& Study Aid via Self-creation & \textit{“Writing my own cheat sheet helps me remember and understand.”} (P28) \\
\cline{2-3}
& Personalized Content \& Format & \textit{“I can include information I need and organize it how I understand.”} (P24) \\
\cline{2-3}
& Anxiety Reduction & \textit{“It gives me more control and reduces my stress.”} (P23) \\
\hline
\multirow{4}{=}{\textbf{Attitudes Can Shift}} 
& Realization of Misalignment & \textit{“The sheet only covered one topic... not that useful.”} (P15) \\
\cline{2-3}
& Space Utilization & \textit{“Too much white space, I could’ve added more.”} (P22) \\
\cline{2-3}
& Efficiency Doubts & \textit{“I felt better equipped to make my own.”} (P33) \\
\cline{2-3}
& Curiosity-Driven Switch & \textit{“I’ve never tried an instructor sheet, I wanted to see what it was like.”} (P4) \\
\hline
\multirow{5}{=}{\textbf{Constraints}} 
& Font Size and Legibility & \textit{“I wrote very small... it was hard to read.”} (P2) \\
\cline{2-3}
& Overchecking & \textit{“I double checked my answers with the cheat sheet even if I was sure my answer was correct and that wasted a bit of my time.”} (P35) \\
\cline{2-3}
& Information Density Trade-offs & \textit{“This took a lot of time and made me question constantly if I was taking in information that would be relevant to the final.”} (P14)\\
\cline{2-3}
& Exam Length vs. Cheat Sheet Use & \textit{“Exam was very long so towards the end I barely looked at my cheatsheet.”} (P9) \\
\cline{2-3}
& Layout Efficiency & \textit{“It was kind of difficult to efficiently find the required information due to how linear and crammed the content was.”} (P21)\\
\hline
\end{tabular}
\end{table*}

\section{Qualitative Results}
% All qualitative codes are summarized in Table \ref{tab:qual_themes}. 
Table \ref{tab:qual_themes} summarizes the qualitative themes and includes selected representative quotes from participants.
We organize our findings into three categories. First, we describe the rationales behind their cheat sheet choices: how they evaluated different cheat sheet formats and what values, such as control, trust, and perceived fairness, shaped their reasoning. 
Second, we explore how these attitudes shifted over time, as students reflected on the outcomes of their initial choices and adapted their preparation strategies. 
Third, we examine the practical challenges students faced in using their cheat sheets, such as navigating space limitations, font size and legibility issues, and balancing information density with usability.

We observed tensions between values and realities: for instance, students who initially distrusted the instructor-provided cheat sheet sometimes acknowledged its exam alignment, while others who preferred personalization found it cognitively burdensome to condense material under pressure. These findings reflect a broader question: what does ``support'' mean when students are given a choice, and how do they interpret pedagogical intent through the tools provided?

\subsection{Rationales Behind Choices}
% \subsubsection{Trust in Expertise and Efficiency.}
\subsubsection{Rationales for instructor-provided cheat sheets}
Students who selected the instructor-provided cheat sheet often viewed it as a more efficient and trustworthy option. Rather than framing this choice as passive or disengaged, participants described strategic reasoning rooted in beliefs about instructional authority, relevance, and exam alignment.

% \subsubsection*{\textcolor{red}{Trust in Instructor}.} 
Several participants emphasized their trust in the instructor’s experience to anticipate what mattered most. \textit{“I feel like the instructor will include more relevant information to the exam than I would,”} P1 said. Others echoed this logic, highlighting instructors' proximity to assessment design: \textit{“I trust the instructor would make it useful”} (P20) and \textit{“the instructor is in a better position to provide a cheat sheet as they know what’s on the exam”} (P33).

% \subsubsection*{\textcolor{red}{Cognitive Efficiency}.} 
Cognitive efficiency also played a central role in students’ decisions. Preparing a cheat sheet from scratch was seen by many as time-consuming, especially under pressure. P10 explained, \textit{“I can use my study time to focus on reviewing course content rather than making a cheat sheet.”} Similarly, P7 decided to \textit{“prepare as if the cheat sheet didn’t exist, and treat the instructor-provided one as a bonus.”}

% \subsubsection*{\textcolor{red}{Exam Alignment}.} 
Others were wary of the risks of creating a suboptimal resource. \textit{“Sometimes I spend a lot of time preparing my cheat sheet, and not studying as broadly,”} P16 reflected. \textit{“Then the topics I focused on were not the main topics covered in the exam.”} For these students, delegating the task to the instructor was not about laziness; it was a form of cognitive off-loading to avoid narrowing their review.

Some participants also appreciated the way a standardized sheet could reduce inequity and pressure. As P48 put it, \textit{“using an instructor-provided cheat sheet keeps the playing field fair for everyone.”} For P17 and P37, it alleviated stress and offered a reprieve from the mental burden of preparation.

% \subsubsection{Personalization through Preparation.}
\subsubsection{Rationales for self-created cheat sheets}
Students who opted to create their own cheat sheets often viewed the process not just as a way to collect information, but as a deliberate method of studying. For many, the act of writing was itself a learning strategy. As P28 explained, \textit{“writing a cheat sheet helps me to understand the concept rather than just memorizing them for the purpose of the exam.”} This reflection was echoed by others who saw cheat sheet creation as a means of active recall and synthesis. P27 simply noted, \textit{“helps me remember the content better as I am making the cheat sheet,”} while P5 framed it as a way \textit{“to study effectively while preparing.”}

% \subsubsection*{\textcolor{red}{Personalized Content}.}
Beyond cognitive reinforcement, personalization was a recurring theme. Students expressed a desire to tailor their sheets to their individual weaknesses in understanding, rather than rely on a generalized resource. P23 described this as a strategic trade-off: \textit{“by choosing self-prepared cheat sheet, this gives me more incentive to study and review notes... alleviating the stresses of an instructor-provided cheat sheet where certain notes may not be there or be hard to find during the midterm.”} Similarly, P12 emphasized the benefit of tailoring content \textit{“based on my knowledge gaps,”} while P34 preferred to \textit{“focus on material that confused me more.”}

% \subsubsection*{\textcolor{red}{Format}.}
Control, both over content and layout, was an equally important motivation. Several students described a discomfort with relying on a cheat sheet they did not assemble themselves. P36 elaborated: \textit{“I don’t like the idea of not knowing what resource / information I will have for the midterm... This won’t be an issue if I just make my own cheat sheet.”} Others pointed to a spatial familiarity that developed through the act of creation. P31 explained, \textit{“I would know where in the sheet I can look to find the relevant material,”} and P42 added, \textit{“I know exactly where to look in my cheat sheet for the question that I need to answer.”}

% \subsubsection*{\textcolor{red}{Anxiety Reduction}.} 
For some, creating a cheat sheet also meant asserting ownership over exam preparation. This choice helped reduce stress (P35), support planning (P53), and ensure alignment with their study priorities (P6, P25, P47). P44 described this as a personal connection to their study process: \textit{“sometimes when looking at the cheatsheet I can understand my own thinking better... I can ‘revisit’ the time when I was writing the notes, making it better for information recall.”}

These reflections show that the choice to self-prepare a cheat sheet was not simply about distrust or dissatisfaction with the instructor’s sheet. Instead, it represented an intentional decision to align study materials with personal learning preferences, prioritizing autonomy, customization, and cognitive engagement.

\subsection{Attitudes Can Shift}
While many students maintained a consistent preference for instructor-provided or self-created cheat sheets, several participants reported changing their stance over the term. 

% \subsubsection{\textcolor{red}{Realization of Misalignment, Space Utilization, and Efficiency Doubts.}}
Some students who initially used the instructor-provided cheat sheet later chose to create their own for the final exam. These shifts were primarily driven by frustration over the limitations of the instructor's version (P15, P22, P46).

\begin{quote}
\textit{Given how limited the information was for the midterm cheatsheet, I would like to create my own for the final... There was too much space left unused and the sheet basically only covered one topic (which was also the easiest topic in my opinion).} (P15)
\end{quote}

These students did not reject the idea of instructor-provided cheat sheets entirely. Rather, their comments suggest a desire for a version that better reflected their knowledge gaps, exam expectations, and need for more efficient layout design. For example, P30 suggested shifting the emphasis from formulas (which are easy to memorize) to definitions (which are harder to recall but more conceptually important).

\begin{quote}
\textit{OCL operation is not hard to memorize. Those definitions are. Thus, the cheat sheet is not very helpful.} (P30)
\end{quote}

% \subsubsection{\textcolor{red}{Curiosity-Driven Switch.}}
Conversely, one participant shifted from self-prepared to instructor-provided, motivated by curiosity and a desire to experiment with unfamiliar formats.

\begin{quote}
\textit{I have never used an instructor provided cheatsheet before, and would like to try it out.} (P4)
\end{quote}

Interestingly, even in switching strategies, P4 offered a reflective critique of their previous approach:

\begin{quote}
\textit{I would be more organized in the layout of my cheatsheet.} (P4)
\end{quote}

\subsection{Constraints}
% \subsubsection{textcolor{red}{Font Size and Legibility.}}
While students appreciated the opportunity to use cheat sheets, their reflections revealed a complex landscape of logistical, cognitive, and affective trade-offs. One recurring theme was space and legibility constraints. Students who opted to include as much information as possible reported difficulties with reading and organization. As P2 put it, \textit{“I wrote very small to fit all the content so sometimes it was hard to read (but still legible).”} P36 echoed, \textit{“I used a light pink which when printed tiny against a white background makes it really hard to read.”} For several students, the physical limits of a double-sided page became a source of frustration, particularly in the final exam: \textit{“I added A LOT of info on my cheat sheet so it would've been nice if we could've had 2 pages double sided since final covered double the content then the midterm”} (P13). 

% \subsubsection{textcolor{red}{Overchecking and Layout Efficiency.}}
A second challenge was time management and over-reliance. Some students found themselves constantly verifying even what they already knew, which paradoxically slowed their progress: \textit{“I felt that I would refer to the cheat sheet a lot rather than trusting what I thought the answer was… I feel like this slowed me down quite a bit compared to an exam without a cheat sheet”} (P23). Others noted that simply locating information was time-consuming: \textit{“It was kind of difficult to efficiently find the required information due to how linear and crammed the content was”} (P21).

Third, students grappled with content selection uncertainty, especially those who created their own sheets. The act of choosing what to include was sometimes anxiety-inducing: \textit{“There was so much information and I wasn't sure what to prioritize putting in my cheat sheet”} (P42). Some expressed regret over missing key definitions, examples, or contextual clues: \textit{“My cheat sheet didn’t have some definitions or key terms in context that was on the exam”} (P21), and \textit{“I wish I added more examples”} (P39).

Lastly, a few critiques addressed the instructor-provided sheet specifically. These centered on perceived mismatches with the actual exam: \textit{“A lot of the information on the instructor-provided sheet was kinda not needed, so sorting through it was rough”} (P48).  P35 went further, questioning the overall benefit: \textit{“Having the cheat sheet did not improve my performance as none of the questions were concept-based.”}

These reflections suggest that cheat sheets are not a panacea for exam anxiety or performance issues. Instead, they function as tools that students must actively adapt to their own learning strategies, often negotiating trade-offs between completeness and clarity, certainty and efficiency, and support and independence.

\section{Quantitative Results}

% \outline{explain any data gathering issues (like logging errors), special post-processing steps, and outlier removal.}

% For each combination of participant, \f{technique}, and \f{task}, trial times more than 3 standard deviations from the  mean time were excluded as outliers. In total, 405 trials (1.3\%) were removed.

% \outline{summarise analysis method}

\begin{figure*}[tb]
	\centering
	\includegraphics[width=\twocolwidth]{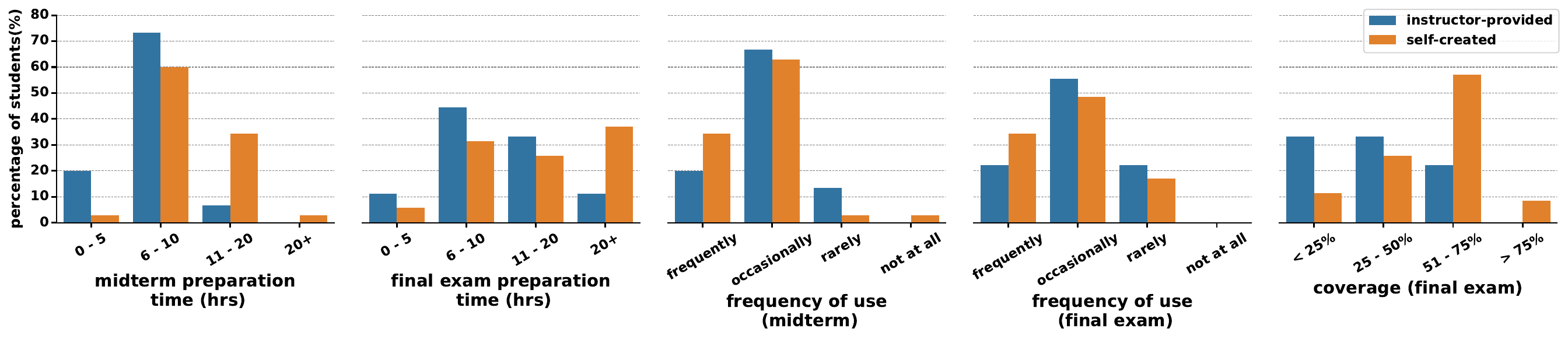}
	\caption{Distribution of preparation time, frequency of cheat sheet use, and final exam coverage by cheat sheet type.}
	\label{fig:combine_vis} 
	% use this common caption style for all results figures: <name of DV> by <factors on x-axis> for each <factors using colour given in legend>
	% in at least first figure, make it clear what your error bars are
	% make sure you reference this figure in the task text
\end{figure*}

% In the analysis to follow, a \f{technique} \by{} \f{task} \by{} \f{block} ANOVA with Tukey HSD post hoc tests was used, unless noted otherwise. When the assumption of sphericity was violated, degrees of freedom were corrected using Greenhouse-Geisser ($\epsilon < 0.75$) or Huynh-Feldt ($\epsilon \geq 0.75$).
% Residuals for \m{Selection Time} were not normally distributed, so log transformed values were used for statistical analysis.
% For each measures, trials were aggregated by participant and factors being analysed.
% In the analysis that follows, we examined whether the distribution of preparation time (0-5 hours, 6-10 hours, 11-20 hours, more than 20 hours) differed significantly between participants who used instructor-provided cheat sheets and those who created their own. Given that preparation time is an ordinal variable, we used Chi-square tests of independence to compare the frequency distributions across cheat sheet types. Residuals and assumptions were not evaluated since the Chi-square test does not require normality or homoscedasticity of data. 
Participants were asked about their preparation time, cheat sheet usage patterns, and the perceived impact of cheat sheets on their final exam performance, as shown in Figure \ref{fig:combine_vis}. Across the two exams, most students preferred to create their own cheat sheets (70\% vs. 30\% for the midterm; 79.6\% vs. 20.4\% for the final). These proportions provide context for the quantitative comparisons that follow.

\textbf{Finding 1: Exam preparation time varied by cheat sheet type for the midterm but not for the final exam.}
We examined whether the distribution of preparation time differed significantly between participants who used instructor-provided cheat sheets and those who created their own. Given that preparation time is an ordinal variable, we used Chi-square tests of independence to compare the frequency distributions across cheat sheet types.  
For the midterm, participants using self-created cheat sheets reported much longer preparation time than those using instructor-provided cheat sheets (\(
\chi^2 = 12.58, \quad p = 0.0056 < 0.05, \quad df = 3
\)). However, for the final exam, this difference was not statistically significant (\(\chi^2 = 2.34, \quad p = 0.51, \quad df = 3
\)). One possible explanation is that as exam content becomes more complex, instructor-provided cheat sheets no longer reduce preparation time, as students still need to invest effort to fully understand and integrate the material.

\textbf{Finding 2: Over half of the participants got used to referring to the cheat sheet occasionally during exams.}
When asked how frequently they consulted the cheat sheet, the majority of participants said they “occasionally” used it, with around 64\% for the midterm and 50\% for the final exam. In both exams, frequent use was reported by about 30\% of participants, while only a small fraction (less than 20\%) stated that they “rarely” or “never” referred to the cheat sheet. This pattern indicates how cheat sheets served more as a backup reference than as a central tool for most participants.

\textbf{Finding 3: Students using self-created cheat sheets reported greater perceived coverage of the information they needed on the final exam.}
When participants were asked how much of the final exam information was covered by their cheat sheets, a key difference emerged: self-created cheat sheets were more likely to cover a larger portion of the exam material. Specifically, 57\% of participants using self-created cheat sheets reported coverage of 51–75\% of the exam content, while instructor-provided cheat sheets were more evenly distributed across lower coverage levels.
% This suggests that students who created their own cheat sheets may have included more comprehensive or tailored content compared to the instructor-provided versions. Compared to instructor-provided versions, self-created sheets may have better supported active and strategic engagement with the material.

\textbf{Finding 4: Consulting the cheat sheet while working through the exam was the most commonly selected usage strategy.}
The most popular strategy for using the cheat sheet was to “refer to the cheat sheet continuously as I worked through the exam,” reported by 73\% of participants. Other strategies, such as “skimming the cheat sheet at the beginning of the exam” (23\%) and “only referring for questions I was unsure about” (13\%), were much less common. This shows that many participants viewed the cheat sheet as an ongoing aid rather than a one-time reference tool.

\textbf{Finding 5: Cheat sheets boosted participants’ confidence and efficiency.}
Participants reported positive effects of cheat sheets on their time management. A majority (55\%) felt that cheat sheets “helped me work faster and more confidently,” and 57\% said cheat sheets “helped reduce time spent recalling definitions or formulas.” Only a small proportion (11\%) reported that cheat sheets slowed them down because they spent time searching for information.

\begin{figure}[h]
	\centering
	\includegraphics[width=\onecolwidth]{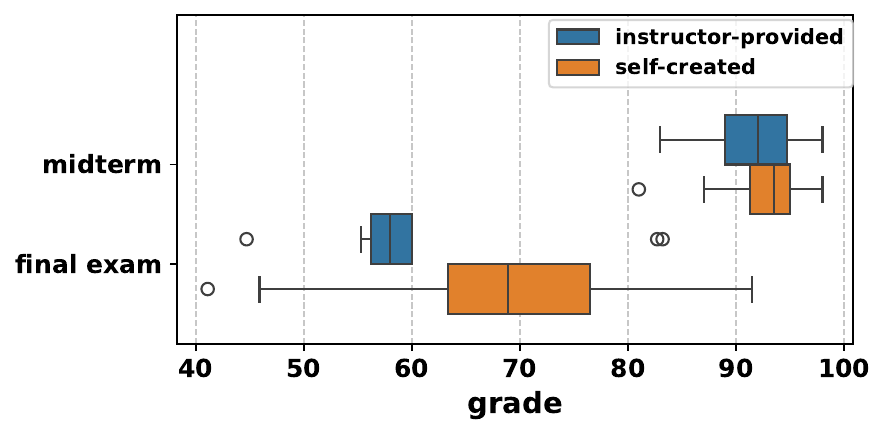}
	\caption{Grade distributions for midterm and final exams by cheat sheet type.}
	\label{fig:t_vis} 
	% use this common caption style for all results figures: <name of DV> by <factors on x-axis> for each <factors using colour given in legend>
	% in at least first figure, make it clear what your error bars are
	% make sure you reference this figure in the task text
\end{figure}

\textbf{Finding 6: Cheat sheet type was not associated with performance outcomes, although a slight trend was observed in the final exam.}
To examine whether using instructor-provided or self-created cheat sheets was linked to exam performance, we conducted independent samples t-tests for midterm and final exams.
For the midterm, there was no significant difference in performance between instructor-provided cheat sheet users (mean = 91.7) and self-created cheat sheet users (mean = 92.7), where \( t = -0.76, p = 0.46, Cohen's ~d = -0.25 \). For the final exam, while not statistically significant, a trend emerged showing higher average scores for self-created cheat sheet users (mean = 69.5) compared to those who used instructor-provided cheat sheets (mean = 61.8), where \( t = -1.65, p = 0.13, Cohen's ~d = -0.63 \). Although this difference did not reach the conventional threshold for statistical significance, the moderate effect size suggests that self-created cheat sheets may have been more beneficial in supporting final exam performance, as shown in Figure \ref{fig:t_vis}.

\section{DISCUSSION}
% There is a quiet but significant tension in how cheat sheets are understood in our study: either as a supportive tool designed to scaffold learning or as a potential shortcut that risks undermining critical engagement with surface-level review. Our findings highlight how cheat sheets occupy a contested space between instructor intentions and student interpretations, revealing deeper questions about trust, agency, and the meaning of support in a learning environment. 
% In what follows, 
We first revisit the research questions by synthesizing the main findings across students’ choices, preparation strategies, usage patterns, and exam performance. We then contextualize these findings with the design rationale for the instructor-provided cheat sheet and discuss implications for cheat sheet policy and design.

\subsection{Revisiting the Research Questions}
With respect to RQ1, the factors shaping students’ preferences went beyond convenience alone. Students who preferred instructor-provided cheat sheets often valued trust in instructor expertise, perceived alignment with the exam, and the efficiency of delegating part of the preparation process. By contrast, students who preferred self-created cheat sheets often valued personalization, control over content and layout, and the act of creating the sheet as a form of studying. These preferences suggest that students were not simply deciding between doing more work and doing less work, but between different forms of support: one grounded in trusted guidance and reduced uncertainty, the other in autonomy, ownership, and active engagement. At the same time, these choices may also have been shaped by factors not directly measured in our study, such as students’ confidence in their own notes, their level of engagement with lectures, or their reliance on instructor-curated materials during preparation.

% In terms of RQ2, students’ cheat sheet choices related to preparation, use, and performance in ways that were meaningful but not straightforwardly hierarchical. Students who created their own cheat sheets reported spending significantly more time preparing for the midterm, and they were more likely to perceive their sheets as covering a larger portion of the exam material. Their qualitative accounts suggest that this was not merely extra labour, but part of how they studied: constructing the sheet helped them synthesize content, identify knowledge gaps, and build familiarity with the layout of the material they would later reference. Students who chose instructor-provided cheat sheets, by contrast, often described that option as a way to offload part of the preparation process and focus on reviewing course content more broadly. Across both formats, cheat sheets were commonly used as an ongoing aid during the exam, and many students reported that the sheets improved their confidence and efficiency. At the same time, we did not observe statistically significant performance differences by cheat sheet type, although the final exam showed a moderate, non-significant trend favouring self-created cheat sheets.
For RQ2, students’ choices of cheat sheet related to their preparation strategies, usage patterns, and performance in different ways. In terms of preparation strategies, students who created their own cheat sheets reported significantly longer preparation time for the midterm, though this difference did not remain for the final exam. Qualitative accounts further suggest that self-created cheat sheets were often treated as part of the studying process itself, whereas instructor-provided cheat sheets were often valued as a way to reduce preparation burden and off-load decisions about what content to include. For usage patterns, students across both formats commonly described cheat sheets as an ongoing aid during the exam. Many participants also reported that cheat sheets improved their confidence and efficiency by reducing time spent recalling formulas or definitions. At the same time, qualitative reflections showed that usage was not always seamless: some students reported over-reliance, difficulty locating information quickly, or frustration with legibility and space constraints. In terms of performance, we did not observe statistically significant differences between instructor-provided and self-created cheat sheet users on either the midterm or the final exam. Although the final exam showed a moderate, non-significant trend favouring self-created cheat sheets, the overall results do not provide strong evidence that one cheat sheet format was consistently superior in supporting exam performance.
Overall, these findings suggest that students’ cheat sheet choices were related to how they prepared and how they used the resource during the exam, but less clearly related to measurable performance outcomes. Rather than identifying a single best format, the results point to different trade-offs in how instructor-provided and self-created cheat sheets support students’ preparation and exam-taking practices.

\subsection{Design Rationale for the Instructor-Provided Cheat Sheet}
The instructor shared that the instructor-provided cheat sheet was never meant to offer direct answers to exam questions. It was an attempt to provide students with a scaffold, such as a curated set of Object Constraint Language (OCL) notations and flowcharts, intended to promote syntactic precision and structured thinking. \textit{“I wanted them to navigate the material, not memorize it,”} they explained. Yet this scaffolding was, by design, not refined or optimized for readability. It replicated the slides from class, with minimal editing, to ensure consistency and transparency, even if that meant students needed to invest time locating what was relevant. In this friction between ease of use and fidelity to course content lay the instructor’s belief that genuine learning emerges from grappling with messy, imperfect materials. For the instructor, the cheat sheet’s limitations were not a failing to be corrected but an intentional prompt: a way to remind students that learning is about navigating uncertainty, not circumventing it. They viewed the cheat sheet as a supplementary resource, one that asked students to engage in selection and prioritization, resisting the temptation to see it as an answer key.

\subsection{Rethinking Cheat Sheet Design}
Although no statistically significant differences in exam performance were observed between the two formats, our findings show that cheat sheets function less as performance aids and more as pedagogical artifacts that reflect how students engage with course content. Based on this, we suggest that instructors may consider encouraging students to create their own cheat sheets, rather than relying on instructor-provided versions. This aligns with learner-centered teaching approaches, which emphasize the role of student autonomy in constructing meaningful learning experiences \cite{kerimbayev2023student, zimmerman2002becoming}.

While concerns about fairness and academic integrity are often cited as reasons to restrict the use of self-created cheat sheets \cite{cooksey2019handle, singh2025promoting}, our findings show that these concerns may be overstated. 
The structural constraints imposed by the exam, such as limiting students to a double-sided A4 paper, cap what can be included. Within these boundaries, self-created cheat sheets allow students to engage with the material more deliberately, allowing them to tailor their sheets to their individual learning styles and areas of uncertainty. This process reflects a shift toward personalization and strategic learning, where learners are active participants in shaping how they study \cite{pintrich2002role}.
The act of constructing a cheat sheet, especially by hand, requires students to prioritize, organize, and synthesize information. Unlike digital copy-and-paste habits that may encourage passive review, handwriting has been shown to improve retention and conceptual understanding by requiring cognitive reprocessing of material \cite{mueller2014pen}. Many participants in our study echoed this: they described how making their own cheat sheet helped them \textit{“remember the content better”} (P27) or provided a familiar spatial layout they could rely on during the exam (P31, P42). These benefits suggest that cheat sheet creation functions not as a shortcut, but as a form of self-regulated learning \cite{zimmerman2002becoming}.
It is worth noting that the most pronounced quantitative effect we observed concerned preparation behaviour: students who created their own cheat sheets spent significantly more time preparing for the midterm. This additional investment, even without performance gains, highlights how self-creation may cultivate productive study habits and deeper engagement with material.

However, the design of cheat sheet policies must remain inclusive. Students with disabilities or differing learning needs may require alternatives such as typed formats or oral accommodations. Inclusive pedagogy emphasizes the importance of flexibility and equivalence in assessment design \cite{hall2007supporting}, and cheat sheet policies should reflect this by allowing structurally different but pedagogically equivalent options.

% We suggest that cheat sheets should be treated as intentional components of the learning process, not as logistical tools or exam loopholes. When students are invited to construct their own supports within fair and inclusive boundaries, they not only deepen their understanding but also develop transferable skills in information prioritization, critical evaluation, and exam preparation. Instructors who treat cheat sheet creation as a structured learning activity, not just a permission, can help students move from passive consumers of content to active stewards of their own learning. This framing aligns with \citet{https://doi.org/10.1002/j.2168-9830.2004.tb00809.x}'s account of active learning, which emphasizes meaningfuk  

We suggest that cheat sheets should be treated as intentional components of the learning process, rather than merely as logistical tools or exam loopholes. When students are invited to construct their own supports within fair and inclusive boundaries, the process may deepen their engagement with course content while helping them develop transferable skills in information prioritization, organization, and exam preparation. By framing cheat sheet creation as a structured learning activity, rather than simply permitting students to bring notes, instructors can position students as active participants in their own learning. This framing aligns with \citet{https://doi.org/10.1002/j.2168-9830.2004.tb00809.x}’s account of active learning, in which students engage in meaningful activities and think about what they are doing; in this context, selecting, organizing, and synthesizing material become part of the learning process itself. 

\section{Limitations and Future Work}
Our study has several limitations, each of which opens up a corresponding direction for future work.
\subsubsection*{Course context.} This study was conducted within a single term of a senior-level undergraduate software requirements course, under a specific instructor and cheat sheet policy. While this allowed us to study students’ choices in an authentic setting, those choices may look different in other computing courses, especially introductory programming courses, courses with different exam structures, or courses delivered online, asynchronously, or in hybrid formats. Future work could examine how students’ choices and reasoning shift across these settings.

\subsubsection*{Artifact-level analysis of cheat sheets.}
Our study focused on how students reasoned about the available options, rather than on a direct comparison of the artifacts they used. As a result, we cannot say with certainty how much instructor-provided and self-created cheat sheets overlapped in content, organization, information density, or exam alignment. Nor can we separate students’ reactions to the instructor-provided format from their reactions to the particular implementation of that sheet in this course. Future work could compare the artifacts directly and examine how those material differences shape students’ preferences and exam preparation.

\subsubsection*{AI in exam preparation.} Our study did not examine AI-supported cheat sheet creation, and students in our data did not identify AI as a reason for choosing one cheat sheet format over the other. This remains an important direction for future work, but not simply because AI may make cheat sheet preparation easier. In our setting, the value of self-created cheat sheets appeared to lie partly in the act of selecting, organizing, and compressing course content into a personally meaningful resource. AI support may alter this process substantially: while it may help summarize or reformat large volumes of material \cite{Sorescu2025DevelopingITA}, it may also reduce the cognitive work involved in constructing the sheet and blur the distinction between self-created and externally generated supports. Moreover, AI may not reliably capture the instructor’s pedagogical emphasis, especially in courses where key priorities are conveyed through lectures, examples, or repeated in-class framing rather than slides alone. Future work should examine not only whether students use AI to prepare cheat sheets, but also how AI-mediated preparation changes the learning value of the artifact itself.

\section{CONCLUSION}
In this work, we have explored what factors shaped students’ preferences between instructor-provided and self-created cheat sheets, as well as how students’ choices of cheat sheets related to their performance, preparation strategies, and usage patterns. Our findings highlight how these preferences and strategies reflect students’ broader approaches to learning and their sense of agency in navigating course content. 
% Future work could compare the actual contents of instructor-provided and self-created cheat sheets to better understand how these differences influence study approaches and learning outcomes. 
We hope that this work can inform practices and policies around cheat sheet use in exams, framing them not as static tools, but as dynamic supports that foster individualized learning.

% \input{2-background}

% \input{3-technique}

% \input{4.1-exp-protocol}

% \input{4.2-exp-results}

% \input{5-discussion}

% \input{6-conclusion}

% --------------------------------

%% Acknowledgements 
%% ASK DAN WHICH ONES TO USE:
\begin{acks}
We thank Dr. Maura R. Grossman, Dr. Matthew Lakier, and Dr. Stacey Watson for providing valuable feedback on earlier versions of this work. We also thank the anonymous reviewers for their constructive comments, which helped strengthen the paper, and our participants for sharing their time and insights.
\end{acks}

%% reference section
\bibliographystyle{lib-acm/ACM-Reference-Format}
\bibliography{_references.bib}

% SUBMITTING TO ARXIV
% If you are submitting your paper to arXiv, change the bibliography to be "main_acm.bib" (the same name as your main.tex file) and use the Submit feature in OL to compile the main_acm.bbl file (select arXiv as the submission type). To avoid processing errors on arXiv with your references, your bibliography needs to have the same name as your main.tex file before you compile it.

% Use the arxiv latex cleaner (https://github.com/google-research/arxiv-latex-cleaner) to clean the source before you upload it to arXiv. This is also useful when you upload the source to PCS:
%  arxiv_latex_cleaner <folder> --keep_bib --commands_to_delete <commands, e.g., dv> 
%  You can add "--commands_only_to_delete rev" to your arXiv latex cleaner command to delete rev commands, but keep the rev-text (unlike --commands_to_delete which deletes the inner text too)
% The \Description tag messes up the arXiv HTML. You might want to delete those too

% After cleaning, you NEED to copy the lib-acm folder to the cleaned version. I think the new ACM format messes up the compilation, because they aren't automatically included post cleaning.

%% appendices
%% If your work has an appendix, this is the place to put it. The TC: comments tell the word count scripts to ignore appendix.  

%TC:ignore  
\appendix
\makeatother
\clearpage
% Reset the figure count and add an A prefix to distinguish it from your other figures
\renewcommand\thefigure{\thesection.\arabic{figure}}
\renewcommand\thetable{\thesection.\arabic{table}}
\setcounter{figure}{0}
\setcounter{table}{0}
\section{Survey Questions}
\label{apx:survey questions}
\subsection{Survey 1}
\begin{enumerate}[label=(\arabic*)]
\item Please select the gender identity option(s) with which you identify. Select ALL that apply
\begin{enumerate}[label=\Alph*.]
\item Agender
\item Gender non-conforming
\item Man (includes cis men, trans men, and anyone who identifies as a man)
\item Non-binary
\item Questioning
\item Trans
\item Women (includes cis women, trans women, and anyone who identifies as a woman)
\item Another gender identity
\item I prefer not to answer
\end{enumerate}

\item What is your first language?
\begin{enumerate}[label=\Alph*.]
\item English
\item Other (please specify)
\end{enumerate}
% \item If you selected "Other" for your first language, would you mind sharing the First languages you speak?

\item What is your year of study?
\begin{enumerate}[label=\Alph*.]
\item First year
\item Second year
\item Third year
\item Fourth year and above (undergraduate level)
\item I am a graduate student
\end{enumerate}

\item Have you used cheat sheets in exams before? (A \textbf{cheat sheet} (also cheatsheet) is a concise set of notes used for quick reference during exams.)
\begin{enumerate}[label=\Alph*.]
\item Yes
\item No
\end{enumerate}

\item What is your typical method for exam preparation? \textit{Select ALL that apply.}
\begin{enumerate}[label=\Alph*.]
\item Reviewing notes
\item Practicing problems
\item Creating study aids
\item Other (please specify)
\end{enumerate}

\item Which type of cheat sheet would you prefer to use for the midterm?
\begin{enumerate}[label=\Alph*.]
\item Instructor-provided
\item Self-prepared
\end{enumerate}

\item Do you think using a cheat sheet will reduce your exam stress?
\begin{enumerate}[label=\Alph*.]
\item Yes
\item No
\item Not sure
\end{enumerate}

\item Why did you choose this type of cheat sheet for the midterm? (Open-ended Question)
\end{enumerate}

\subsection{Survey 2}
\begin{enumerate}[label=(\arabic*)]
\item How much time did you spend preparing for the midterm?
\begin{enumerate}[label=\Alph*.]
\item 0 - 5 hours
\item 6 - 10 hours
\item 11 - 20 hours
\item More than 20 hours
\end{enumerate}

\item Which type of cheat sheet did you use for the midterm?
\begin{enumerate}[label=\Alph*.]
\item Instructor-provided
\item Self-prepared
\end{enumerate}

\item Did the cheat sheet contain all the information you needed for this midterm?
\begin{enumerate}[label=\Alph*.]
\item Yes
\item No
\end{enumerate}

\item How often did you refer to the cheat sheet during the exam?
\begin{enumerate}[label=\Alph*.]
\item Frequently
\item Occasionally
\item Rarely
\item Not at all
\end{enumerate}

\item Do you think using the cheat sheet improved your performance?
\begin{enumerate}[label=\Alph*.]
\item Yes
\item No
\end{enumerate}

\item How did the cheat sheet impact your time management during the exam? (Open-ended question)
\item What challenges, if any, did you face when using the cheat sheet? (Open-ended question)
\item Do you plan to use the same type of cheat sheet for the final exam?
\begin{enumerate}[label=\Alph*.]
\item Yes
\item No
\end{enumerate}

\item If you answered "No" to the previous question, please explain why. (Open-ended question)
\item If you used a self-prepared cheat sheet for this midterm, how would you improve it for future use? If you used an instructor-provided cheat sheet for this midterm, what suggestions do you have for improving it? (Open-ended question)
\end{enumerate}

\subsection{Survey 3}
\begin{enumerate}[label=(\arabic*)]
\item How much time did you spend preparing for the final exam?
\begin{enumerate}[label=\Alph*.]
\item 0 - 5 hours
\item 6 - 10 hours
\item 11 - 20 hours
\item More than 20 hours
\end{enumerate}

\item Which type of cheat sheet did you use for the final exam?
\begin{enumerate}[label=\Alph*.]
\item Instructor-provided
\item Self-prepared
\end{enumerate}

\item Compared to your midterm, did you make any changes to your cheat sheet for the final (e.g., more definitions, better layout, colour coding, added examples, reduced content for clarity)? (Open-ended question)
\item Did the cheat sheet contain all the information you needed for this final exam?
\begin{enumerate}[label=\Alph*.]
\item Yes
\item No
\end{enumerate}

\item Roughly, what percentage of the information you needed for the final exam was covered in your cheat sheet?
\begin{enumerate}[label=\Alph*.]
\item Less than 25\%
\item 25\% - 50\%
\item 51\% - 75\%
\item Greater than 75\%
\end{enumerate}

\item What was your strategy for using the cheat sheet during the exam? \textit{Select ALL that apply}
\begin{enumerate}[label=\Alph*.]
\item I completed all the questions first, and then referred to the cheat sheet for answers I was unsure about.
\item I referred to the cheat sheet continuously as I worked through the exam.
\item I first skimmed the cheat sheet at the beginning of the exam to refresh my memory, then started answering questions.
\item I reviewed the cheat sheet primarily for formulas / definitions / examples when needed.
\item I didn't get a chance to review the cheat sheet during the exam.
\item Other (please specify)
\end{enumerate}

\item How often did you refer to the cheat sheet during the final exam?
\begin{enumerate}[label=\Alph*.]
\item Frequently
\item Occasionally
\item Rarely
\item Not at all
\end{enumerate}

\item How did the cheat sheet impact your time management during the final exam? \textit{Select ALL that apply}
\begin{enumerate}[label=\Alph*.]
\item It helped me work faster and more confidently.
\item It helped reduce time spent recalling definitions or formulas.
\item It slowed me down because I spent time searching for information.
\item It did not significantly affect my time management.
\item I did not use a cheat sheet.
\item Other (please specify)
\end{enumerate}

\item Do you think using the cheat sheet improved your performance?
\begin{enumerate}[label=\Alph*.]
\item Yes
\item No
\end{enumerate}

\item What challenges, if any, did you face when using the cheat sheet? (Open-ended question)
\end{enumerate}

% appendix figures and table should always be "here" layout with [h]
% \begin{figure}[h]
% 	\centering
% 	\includegraphics[width=0.47\textwidth]{figures/tbd.pdf}
% 	\caption{Descriptive figure title: (a) this part of the figure; (b) this part of the figure. Optional other general figure information, like credit, colours, error bars, etc.}
% \label{apx:fig:example}
% \end{figure}

% % appendix figures and table should always be "here" layout with [h]
% \begin{table}[h]
%     \caption{\m{Other Time} ANOVA main effects and interactions }
%     % link to table tex
%     \input{tables/anova-results}
%     \label{tab:othertime-anovas}
% \end{table}

%TC:endignore 

\end{document}